\documentclass[aip,jap,amsmath,amssymb,reprint]{revtex4-1}

\usepackage{graphicx}
\usepackage{dcolumn}
\usepackage{bm}
\usepackage{textcomp}
\usepackage{ulem}
\begin{document}

\preprint{Hoof et al. (2012)}

\title[Enhanced quality factors and force sensitivity for AFM in liquid]{Enhanced quality factors and force sensitivity by attaching magnetic beads \\ to cantilevers for atomic force microscopy in liquid}

\author{Sebastian Hoof}
  \affiliation{London Centre for Nanotechnology, University College London, 17--19 Gordon Street, London WC1H 0AH, United Kingdom}
  \affiliation{Fakult\"at f\"ur Physik und Astronomie, Ruprecht-Karls-Universit\"at Heidelberg, 69120 Heidelberg, Germany}
\author{Nitya Nand Gosvami}
  \altaffiliation[Present address: ]{Department of Mechanical Engineering and Applied Mechanics, University of Pennsylvania, Philadelphia, Pennsylvania 19104, USA}
  \affiliation{Department of Chemistry, University College London, 20 Gordon Street, London, WC1H 0AJ, United Kingdom}
\author{Bart W. Hoogenboom}
 \email{b.hoogenboom@ucl.ac.uk.}
 \affiliation{London Centre for Nanotechnology, University College London, 17--19 Gordon Street, London WC1H 0AH, United Kingdom}
 \affiliation{Department of Physics and Astronomy, University College London, Gower Street, London WC1E 6BT, United Kingdom}

\date{\today}

\begin{abstract}
Dynamic-mode atomic force microscopy (AFM) in liquid remains complicated due to the strong viscous damping of the cantilever resonance. Here we show that a high-quality resonance ($Q>20$) can be achieved in aqueous solution by attaching a microgram-bead at the end of the nanogram-cantilever. The resulting increase in cantilever mass causes the resonance frequency to drop significantly. However, the force sensitivity --- as expressed via the minimum detectable force gradient --- is hardly affected, because of the enhanced quality factor. Via the enhancement of the quality factor, the attached bead also reduces the relative importance of noise in the deflection detector. It can thus yield an improved signal-to-noise ratio when this detector noise is significant. We describe and analyze these effects for a set-up which includes magnetic actuation of the cantilevers and which can be easily implemented in any AFM system that is compatible with an inverted optical microscope.
\end{abstract}



\maketitle

\section{\label{sec:intro}Introduction}
Since its invention, atomic force microscopy (AFM) has become a routine technique for surface characterization at high spatial resolution. Of the various modes of AFM operation, dynamic-mode AFM offers the advantage of reduced drift effect and reduced lateral drag forces on the sample, thus greatly contributing to the versatility of the technique\cite{garcia:review, allison:review}. In dynamic-mode AFM, the cantilever is oscillated at its resonance frequency and the interaction with the sample is probed via changes in the cantilever oscillation. This is significantly more complicated in liquid, because of viscous damping. In aqueous solution, the natural environment for biological molecules, cantilever quality factors are of order one.

Such low quality factors imply a less harmonic and thus harder to interpret cantilever response, yield reduced force sensitivity and, for stiffer cantilevers, an increased relative importance of the noise in the detection of the cantilever deflection. These problems have at least partly been overcome in specialized equipment, by the development of low-noise deflection detectors\cite{hoogenboom:interferometer,fukuma:deflectiondetector} and operation at small oscillation amplitudes, resulting among others in atomic resolution on solid-liquid interfaces by frequency-modulation AFM\cite{fukuma:fm, hoogenboom:fm}. In an alternative, ``diving-bell'' approach, only a small portion of the AFM tip was dipped in the liquid medium\cite{ledue:divingbell}, leading to significantly higher quality factors than when the whole cantilever is submerged. The problems persist, however, in many standard AFM systems.

An even more general problem lies in the actuation of the cantilever resonance. A standard piezo-acoustic drive actuates the whole fluid cell or at best the cantilever support chip, and simultaneously excites a wide range of mechanical resonances, commonly known as the ``forest of peaks'' that is observed in liquid AFM \cite{Schaeffer1996}. Such spurious mechanical resonances usually dominate the low-$Q$ resonance of the cantilever itself, which makes it hard to tune to the cantilever resonance or specifically enhance its effective $Q$ by external positive feedback\cite{tamayo:dfm} and which affects any quantitative interpretation of phase contrast of measured frequency shifts \cite{kokavecz:resonances}. In addition, because of the macroscopic nature of these spurious resonances, they are prone to drift.

These problems have motivated a wide variety of approaches to uniquely actuate the cantilever resonance, i.e., without actuating any other, spurious resonances. The cantilever response should thus closely resemble that of a simple harmonic oscillator. While progress has been made by optimizing piezo-acoustic actuation \cite{maali:piezoimprovement,carrasco:piezoimprovement,asakawa:piezoimprovement}, it has been suggested that the focus in the future should be on ``direct'' actuation methods\cite{xu:forcemod}. Magnetic \cite{han:magnetic,Jarvis:magnetic}, photothermal\cite{Ratcliff1998,Ramos2006} and electrostatic actuation\cite{Umeda2010} provide valuable alternatives that, unlike piezo-acoustic methods, only excite the cantilever itself.

In this work, we have attempted to overcome the problems related to viscous damping by enhancing the (intrinsic) cantilever quality factor. To do so, we deliberately increased the inertial mass of the cantilever by attaching beads with tens of \textmu m diameter and \textmu g masses. While there are models to determine the shift in the resonance frequency after the attachment of beads at a specific position on the cantilever, these models are only valid for a bead mass much smaller than the cantilever mass\cite{dohn:beads}. However, to achieve a significant increase in the quality factor, beads of masses greater than the cantilever mass are needed. We chose magnetic beads to facilitate the comparison between piezo-acoustic actuation and ``ideal'' magnetic actuation of the cantilevers.

In this paper, we first describe the experimental setup for magnetic actuation and bead attachment (Section~\ref{sec:expsetup}). Next (Section~\ref{sec:results}), we demonstrate the effect of the beads on the thermal noise spectrum and resonant response of the cantilever to external actuation, and provide an analysis of the minimum force gradient that can be detected by cantilevers with beads of various sizes. We also include a demonstration of AFM imaging in liquid using this method. We summarize our results in Section~\ref{sec:comment}, before giving further details on the magnetic actuation in Appendix~\ref{sec:solenoid}.  We expect such details to be of wider applicability and use for implemention in any AFM system that is mounted on an inverted optical microscope.

\section{\label{sec:expsetup}Experimental setup}

All experiments and operations were carried out with a NanoWizard I AFM system (JPK Instruments, Berlin, Germany), mounted on an inverted light microscope (IX71, Olympus, Tokio, Japan). Results will be shown/discussed for PPP-NCH and PPP-FM silicon cantilevers (Nanosensors, Neuch\^atel, Switzerland). The nominal dimensions (length $\times$ width $\times$ thickness) are $125 \text{\textmu m} \times 30 \text{\textmu m} \times 4 \text{\textmu m}$ and $225 \text{\textmu m} \times 30 \text{\textmu m} \times 3 \text{\textmu m}$ for the NCH and FM cantilevers, respectively. Cantilever deflections were measured with conventional beam deflection, and the detector was calibrated by pressing the cantilever into a hard substrate by ramping the (calibrated) $z$ scanner. Beads were selected from an isotropic magnetic powder (MQP-S-11-9, Magnequench, now part of: Molycorp, Greenwood Village, Colorado, USA). They were selected for their size as measured on the inverted optical microscope, using a calibrated grid on the output of the camera on the microscope. Bead masses were estimated based on their diameter and the mass density of the bead material.

\begin{figure}
\includegraphics{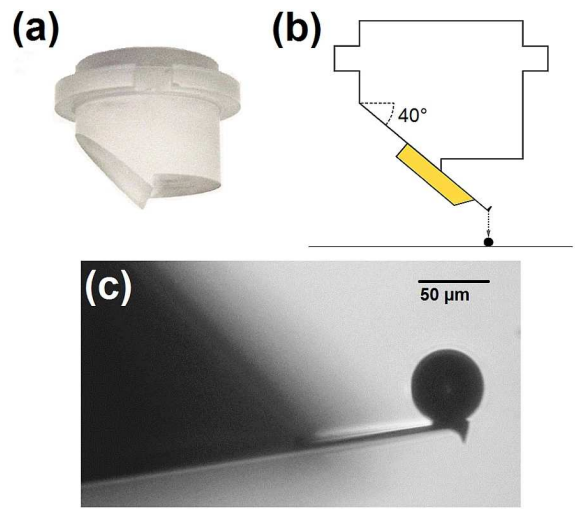}
\caption{\label{fig:bead} (Color online.) (a) Photograph of the home-made cantilever holder designed for attaching beads using the AFM head and the inverted optical microscope. (b) Schematic of the cantilever mounted on the block approaching a bead on a glass slide (not to scale). (c) Optical microscope image of a bead on a PPP-NCH cantilever.}
\end{figure}

Of the various methods to attach particles to cantilevers\cite{gan:beadreview}, we chose the combination of the AFM head and inverted light microscope as a micromanipulator for the attachment process, and used a UV-curing glue (Glass Bond, Loctite) as an adhesive. To optimize the process of bead attachment, a customized, perspex cantilever holder was used, see Fig.~\ref{fig:bead}a-b. It partly follows the design of the standard cantilever holder/fluid cell and is accordingly fixed in the AFM system, but enables us to keep the cantilever at an optimum angle to pick up the adhesive and the bead without risking to touch the underlying glass slide with the support chip. The cantilever was (temporarily) fixed in the customized cantilever holder with Blu-Tack in such a way that the back side of the cantilever faces the sample holder.

To attach the beads, small portions of the adhesive and of the magnetic powder were placed on a glass slide, and the stepper motor of the AFM was used to approach the cantilever to the glass surface. The process was monitored using the optical microscope and the step size of the motor was decreased as the cantilever got closer to the glass slide. After the back side of the cantilever touched the adhesive, the cantilever was retracted. This procedure was repeated for picking up and attaching the bead. It yielded cantilevers with beads as shown in Fig.~\ref{fig:bead}c.

The ideal resonant response of the cantilever was characterized by their thermal noise spectrum. Cantilever resonance frequency $f_0$, quality factor $Q$, and spring constant $k$ were determined from simple harmonic-oscillator fits to the thermal noise power\cite{hoogenboom:nanotech}. For the gold-coated PPP-NCH cantilevers, the microscope had a noise floor of about 0.2~pm/$\sqrt{\text{Hz}}$ for detecting cantilever deflections at frequencies $\lessapprox$70~kHz, and 0.3~pm/$\sqrt{\text{Hz}}$ at frequencies $\gtrapprox$70~kHz, where heterodyne mixing (Heterodyne Thermal Noise Module, JPK Instruments, Berlin, Germany) was required to bring the noise down into the accessible frequency range of the controller and where the noise was increased due to aliasing from higher frequencies. All measurements were carried out in milliQ water unless otherwise specified. Thermal noise curves were recorded before and after attaching a bead at the end of a cantilever. To measure the effect of different beads on the same cantilever, each bead was removed after the measurements by subsequent immersions in isopropanol and water, before drying and attaching another.

For magnetic actuation \cite{florin:magnetic, jayanth:magnetic, malo:magnetic, hofbauer:magnetic}, the beads on the cantilevers were magnetized along their axes by bringing them close to a small, strong (grade: N42) permanent neodymium magnet (F346, Magnet Expert, Sutton-on-Trent, UK). The equivalent field at a distance of 2~mm is approximately 1.4 Tesla\footnote{http://www.kjmagnetics.com/fieldcalculator.asp.}. Compared to other methods of preparing magnetized cantilevers\cite{han:magnetic, revenko:magnetic, xin:comparison, herruzo:comparison, kageshima:magnetic,loke:coatedcl,penedo:magnetostriction}, magnetic beads have the advantage of enabling significant (at least several nm) oscillation amplitudes even for cantilevers with spring constants of $>10$~N/m.

To generate a driving magnetic field, a solenoid was designed, including ferrite core, to fit into one of the objective threads of the inverted optical microscope (Appendix~\ref{sec:solenoid}). This arrangement, with a solenoid under the sample, is similar to the one adopted elsewhere\cite{florin:magnetic, han:magnetic, xin:comparison, malo:magnetic, herruzo:comparison,kageshima:magnetic, hofbauer:magnetic, loke:coatedcl}. Unlike designs where the cantilever is placed inside an electromagnet on the fluid cell\cite{revenko:magnetic, jayanth:magnetic}, this has the advantage of facilitating heat dissipation in the solenoid away from the cantilever and sample. This arrangement implies that AFM imaging with magnetic actuation cannot be performed simultaneously with optical imaging using the inverted microscope, but it is straightforward to switch to optical microscopy in between AFM scans by moving an objective to the optical axis instead of the solenoid.

\section{\label{sec:results}Results and Discussion}
\begin{figure}
\includegraphics{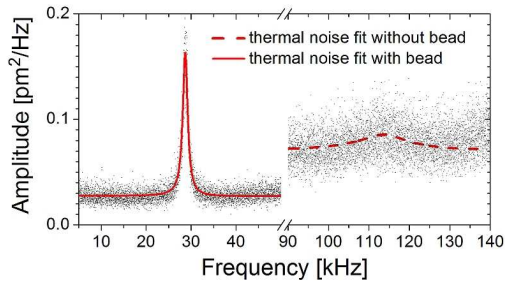}
\caption{\label{fig:comparison2} (Color online.) Thermal noise power spectra in water, for the same PPP-NCH cantilever without (right) and with (left) an attached bead of 55~\textmu m diameter. The solid and dashed lines denote fits with a simple harmonic oscillator model in addition to a constant background noise. The background noise in the right part of the figure is larger because heterodyne mixing was required for the measurement at higher frequencies.}
\end{figure}

Fig.~\ref{fig:comparison2} shows the thermal noise data for a PPP-NCH cantilever without and with a bead of a diameter $d=55$~\textmu m. Not unsurprisingly, we observe a large drop in the cantilever resonance frequency on the attachment of such a large (ca. 0.5~\textmu g) mass to the end of the cantilever (ca. 35~ng, based on the cantilever dimensions and the mass density of silicon). Besides the drop of the resonance frequency, we also find a significant increase of the quality factor. As a consequence of this increase in $Q$, the resonance of the cantilever clearly emerges from the noise of the deflection detector, whereas the resonance is hardly visible without a bead attached to the cantilever (which can only partly be attributed to the higher noise in the upper frequency range, as explained in Section~\ref{sec:expsetup}).

\begin{figure}
\includegraphics{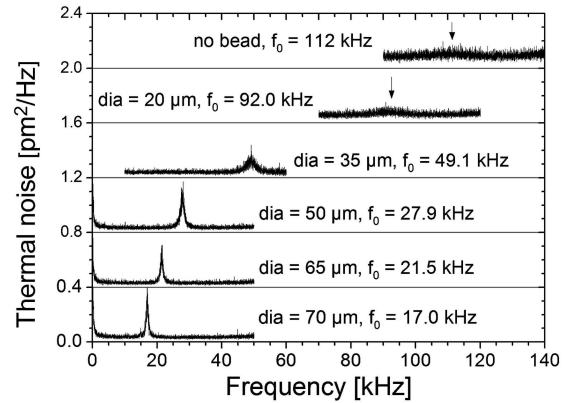}
\caption{\label{fig:beadsize} Thermal noise of cantilevers with beads of different sizes. An offset was added for each set of data for clarity. Horizontal lines denote the zero line for each curve. All measurements were done with the same cantilever except for the bead with $d$ = 20 \textmu m, where another cantilever of the same wafer was used, with a matching resonance frequency (116 kHz) for the unloaded cantilever in water.}
\end{figure}

For a quantification of these effects, this experiment was repeated with different bead sizes, each time recording the thermal noise (Fig.~\ref{fig:beadsize}). $f_0$, $Q$, and $k$ followed from fits to the thermal noise spectrum\cite{hoogenboom:nanotech} and are displayed in Table~\ref{tab:beadsizes}. The bead mass ($m_\text{bead}$) was estimated by assuming them to be perfect spheres and based on a mass density of 7.43 g/cm$^3$ (value provided by the supplier).

\begin{table}
\caption{\label{tab:beadsizes} Effect of beads of different sizes on the dynamic properties of an PPP-NCH cantilever, corresponding to the data displayed in Fig.~\ref{fig:beadsize}.}
\begin{ruledtabular}
\begin{tabular}{ccccc}
$d$ [\textmu m] & $m_\text{bead}$ [\textmu g] & $f_0$ [kHz] & $Q$ & $k$ [N/m]\\
\hline
no bead & - & 112 & \; 7.8 $\pm$ 0.8 & \; 8.7 $\pm$ 2.1\\
20 $\pm$ 5 & 0.03 $\pm$ 0.02 & 92.0 & \; 9.8 $\pm$ 0.6 & 11.2 $\pm$ 1.5\\
35 $\pm$ 5 & 0.17 $\pm$ 0.07 & 49.1 & 16.8 $\pm$ 0.3 & 10.0 $\pm$ 0.4\\
50 $\pm$ 5 & 0.49 $\pm$ 0.15 & 27.9 & 21.1 $\pm$ 0.2 & \; 8.8 $\pm$ 0.2\\
65 $\pm$ 5 & 1.1\; $\pm$ 0.3\; & 21.5 & 23.9 $\pm$ 0.3 & 14.5 $\pm$ 0.3\\
70 $\pm$ 5 & 1.3\; $\pm$ 0.3\; & 17.0 & 21.6 $\pm$ 0.2& 13.0 $\pm$ 0.3\\
\end{tabular}
\end{ruledtabular}
\end{table}

These results confirm the drop in resonance frequency, by up to a factor 5, and the increase in quality factor of nearly a factor 3 for the larger beads, up to values of $Q>20$. This is to be compared to $Q<10$ that is measured here and elsewhere \cite{hoogenboom:fm} for these cantilevers without beads attached. The variations in spring constant can be attributed to the uncertainty in the fits and/or to the adhesive for attaching the different beads.

Similar effects were observed for the softer PPP-FM cantilevers (spring constant for the used unloaded cantilever: $0.97 \pm 0.03$~N/m), where the resonance frequency dropped from 25.8 to 8.11~kHz while the Q-factor increased from $3.51 \pm 0.05$ to $8.79 \pm 0.06$ after attaching a bead with a diameter of 50~\textmu m.

\begin{figure}
\includegraphics{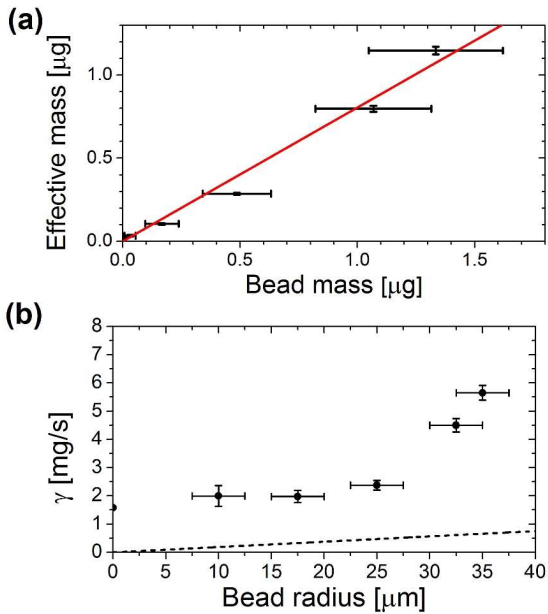}
\caption{\label{fig:effectmass} (Color online.) (a) The effective mass $m^*=k/\omega_0^2$ of the cantilever as a function of the bead mass. The straight line is a linear fit with the constraint of positive values for zero bead mass. (b) Damping coefficient $\gamma=\sqrt{m^* k}/Q$ as a function of bead radius. The dashed line indicates Stokes's law.}
\end{figure}

To further interpret these results, we calculated the effective cantilever mass $m^*=k/\omega_0^2$, which would be the mass in an idealized mass-on-a-spring model of the cantilever (where $\omega_0=2\pi f_0$). Fig.~\ref{fig:effectmass}a shows the $m^*$ as a function of the actual bead mass $m_\text{bead}$. $m^*$ scales linearly  with the bead mass (within the experimental errors). Especially for the larger beads, $m^*\approx m_\text{bead}$, indicating that the dominant contribution to the cantilever inertia is from the bead, and not from the displaced liquid or the cantilever itself. The fact that $m^*<m_\text{bead}$ may be attributed to the fact that the center of the gravity of the bead is not located at the very end of the cantilever. Within the same model, we can assume a viscous damping force of the form $F_v=\gamma v$, with $\gamma=\sqrt{m^* k}/Q$ the damping coefficient and $v$ the velocity of the bead. Plotted as a function of bead radius (Fig.~\ref{fig:effectmass}b), $\gamma$ hardly increases until the bead size becomes significantly larger than the cantilever diameter. This demonstrates that $\gamma$ does not show significant frequency dependence\cite{sader:viscous} over the range explored in our experiments. The results for $\gamma$ can be compared to Stokes's law ($\gamma=6\pi\eta R$, with $\eta$ the viscosity of water), which is valid for a sphere in the limit of a low Reynolds number, where the inertial mass becomes irrelevant.

Higher $Q$ values, as found here, are convenient for finding and tracking the cantilever resonance. The $Q$ value is also an important factor, however, for determining the force sensitivity, or more strictly speaking, the minimum detectable force gradient $\partial F/\partial z |_\text{min}$ in the proximity of the tip to the sample in an AFM experiment. To estimate this quantity, we adopt a commonly used approach\cite{albrecht:highQ, kobayashi:noise}: we focus on elastic interactions only and assume that the force gradient is constant over the whole cantilever oscillation. A force gradient $\partial F/\partial z$ will then cause a well-defined shift $\Delta f/f_0=(2 k)^{-1}\partial F/\partial z$ of the cantilever resonance frequency. Under the assumption that this shift is small compared to the width of the resonance, as is appropriate in liquid, this implies a change in cantilever phase $\Delta\phi=(-Q/k) \partial F/\partial z$ with respect to the phase of a free cantilever oscillation that is driven near the resonance frequency. Since the phase noise of an oscillation is directly related to the amplitude noise $\delta A$ via $\delta\phi=\delta A/A$, the minimum detectable force gradient therefore follows from
\begin{equation}
\left .\frac{\partial F}{\partial z} \right |_\text{min} = \frac{k}{Q} \frac{\delta A}{A} \, ,
\label{eq:minforcegrad}
\end{equation}
with $A$ the amplitude of the oscillation (all amplitudes are rms values unless stated otherwise). In a general AFM experiment, the amplitude noise is the vectorial sum of the thermal cantilever noise  and the amplitude noise caused by the deflection detector, $\delta A=\sqrt{\delta A_\text{the}^2 + \delta A_\text{det}^2}$. If the detector noise is negligible, this can be shown --- via application of the equipartition theorem --- to yield the well-known result\cite{albrecht:highQ, kobayashi:noise}
\begin{equation}
\left .\frac{\partial F}{\partial z} \right |_\text{min} = \frac{1}{A} \sqrt{\frac{4 k k_\text{B} T}{\omega_0 Q}} = \frac{1}{A} \sqrt{4\gamma k_\text{B} T} \, ,
\label{eq:minforcegrad_thermal}
\end{equation}
measured in nN/nm/$\sqrt{\text{Hz}}$, with $\gamma$ the damping coefficient as defined earlier, and where $k_\text{B}$ is the Boltzmann constant and $T$ the absolute temperature. The expression with $\gamma$ is a representation of the fluctuation-dissipation theorem (see, e.g., Ref.~\onlinecite{paul:stochastic}), which states that the thermal force noise power is proportional to the damping $\gamma$ and not affected by the inertial mass $m^*$. Since $\gamma$ does not show significant frequency dependence over the range explored in our experiments (Fig.~\ref{fig:effectmass}b), it follows that the increase in inertial mass of the cantilever does not affect the force sensitivity at resonance.

\begin{figure}
\includegraphics{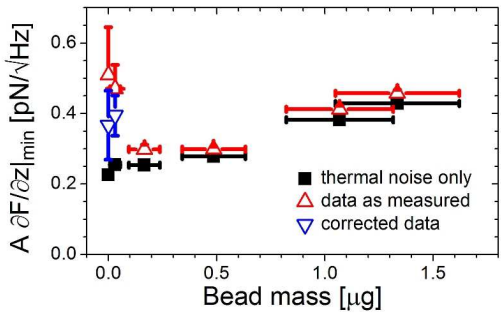}
\caption{\label{fig:minforce} (Color online.) The minimum detectable force gradient for an oscillating cantilever as a function of the mass of an attached bead, multiplied by the oscillation amplitude $A$. Data are shown for the ideal case of negligible detector noise (Eq.~\ref{eq:minforcegrad_thermal}) and for the force gradient based on the measured amplitude noise at resonance (Eq.~\ref{eq:minforcegrad}). In addition, for bead masses $<0.1$~\textmu g, data are also shown based on the amplitude noise that has been corrected for the additional detector noise that was caused by the heterodyne mixing for the measurements at higher frequencies.}
\end{figure}

This can be made more explicit by plotting $A\,\partial F/\partial z |_\text{min}$ in Fig.~\ref{fig:minforce}, both for the ideal case of negligible detector noise (Eq.~\ref{eq:minforcegrad_thermal}, using the measured $k$, $\omega_0$, and $Q$) and based on the measured amplitude noise (Eq.~\ref{eq:minforcegrad}) as observed in Fig.~\ref{fig:beadsize}. The amplitude noise at resonance was obtained from the parameters of the fit of the harmonic oscillator model plus the offset that corresponds to the detector noise. $\partial F/\partial z |_\text{min}$ for any $A$ can be obtained by dividing the values in Fig.~\ref{fig:minforce} by the oscillation amplitude. Without detector noise (Eq.~\ref{eq:minforcegrad_thermal}), we find that the addition of a \textmu g bead to a cantilever of 35~ng hardly affects its force sensitivity. The force sensitivity only shows a significant, though moderate degradation for beads that are significantly larger than the cantilever width (see the increase in $A\,\partial F/\partial z |_\text{min}$ for bead masses $>$1~\textmu g, in Fig.~\ref{fig:minforce}), correlated with the increase in damping in Fig.~\ref{fig:effectmass}b.

\begin{figure}
\includegraphics{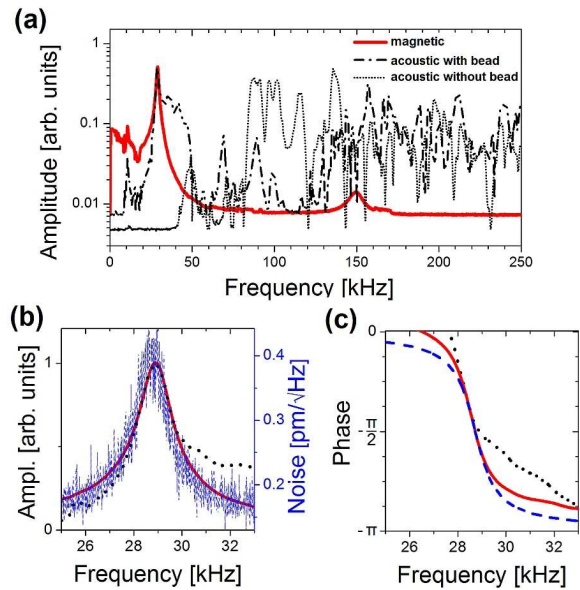}
\caption{\label{fig:comparison} (Color online.) (a) Resonant response of an PPP-NCH cantilever without bead and with bead (diameter ca. 55~\textmu m), compared to magnetic actuation of the same cantilever (with the bead). (b) The response near the resonance, for the cantilever with bead, for both piezo-acoustic (dotted line) and magnetic actuation (solid line), overlayed with the thermal noise (scaled and referred to the right axis). (c) As (b), for the phase response, compared to the theoretical phase response (dashed line) expected from a simple-harmonic-oscillator fit to the thermal noise.}
\end{figure}

In general, Eq.~\ref{eq:minforcegrad_thermal} shows that a drop in resonance frequency will not significantly affect $\partial F/\partial z |_\text{min}$ if it is accompanied by a similarly sized increase in quality factor $Q$ (assuming a constant spring constant $k$). The results in Table~\ref{tab:beadsizes} and Fig.~\ref{fig:minforce} thus demonstrate that the attachment of a larger bead does not significantly affect the sensitivity in an AFM experiment. At the same time, it has the advantage of a sharper and thus more easily detectable cantilever resonance. Importantly, Fig.~\ref{fig:minforce} also shows that the attachment of such beads can {\em improve} the force sensitivity (i.e., minimize $\partial F/\partial z |_\text{min}$) if the detector noise is significant. The attachment of beads can thus be a valuable method to overcome the detector noise in any AFM system, without any adjustments to the microscope itself.  In these experiments, a 50~\textmu m, 0.5~\textmu g bead yields the lowest noise (Fig.~\ref{fig:minforce}) combined with the highest ($> 20$) quality factor (Table~\ref{tab:beadsizes}).

To demonstrate the practical advantage of attaching beads to cantilevers, we actuated the PPP-NCH cantilever without and with bead by standard piezo-acoustic actuation, and compared the resulting resonant response to the result obtained (with bead) by magnetic actuation, see Fig.~\ref{fig:comparison}a. Without bead, the piezo-acoustic actuation yields a broad ``forest of peaks'', as fluid-borne vibrations lead to additional spurious resonances which obscure the cantilever resonance and cause deviations from the theoretically expected frequency response of the cantilever\cite{xin:comparison}. With bead, the spurious resonances are still very noticeable, but the higher $Q$ causes the cantilever resonance to stand out against the spurious background, and at resonance (30~kHz), the response is comparable to the ideal, magnetically actuated response (Fig.~\ref{fig:comparison}b-c). For the magnetic actuation, the fluctuations below 20 kHz are similar to earlier observations\cite{revenko:magnetic} and may be attributed to mechanical vibrations of the coil in this frequency regime.

\begin{figure}
\includegraphics{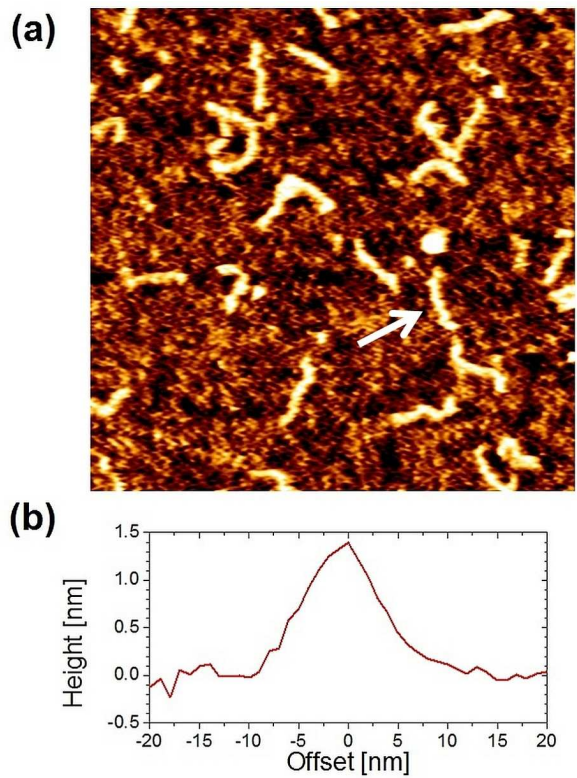}
\caption{\label{fig:imaging} (Color online.) (a) AFM topography of 220 base-pair DNA fragments. The cantilever was driven at its resonance of 29.5~kHz with a peak amplitude of about 2~nm. The amplitude setpoint was optimized for best image contrast and stability. The raw data was modified by first-order line-wise flattening and Gaussian filtering with $\sigma=1.2$~nm (1 pixel). (b) Average cross-sectional profile of the DNA molecule indicated by the arrow.}
\end{figure}

Importantly, the amplitude and phase response around the resonance for both piezo-acoustic and magnetic actuation closely match the ideal response, as determined from the thermal noise spectrum Fig.~\ref{fig:comparison}b-c. We note that this is much less obvious for the first overtone of the cantilever (with bead), which is observed at 150~kHz, such that magnetic actuation would still be preferable if the attached bead is magnetic and a driving field can be generated.

Finally, to illustrate the use of the cantilevers with beads, we imaged double-stranded DNA fragments of 220 base pairs in buffer solution (Fig.~\ref{fig:imaging}), using standard intermittent-contact (''tapping''/ amplitude-modulation) AFM and magnetic actuation of an PPP-NCH cantilever with a 55 \textmu m diameter bead. The DNA fragments were adsorbed on a mica substrate that was pre-treated by 2-minute incubation in 10 mM NiCl$_2$-solution. The DNA was adsorbed (at a concentration of about 10~\textmu g/mL) and imaged in 3 mM NiCl$_2$, 10 mM HEPES pH 7.0 aqueous solution. We found that the used AFM set-up was too sensitive to acoustic and mechanical vibrations\cite{labuda:afmsimulation} for resolving finer details on the DNA\cite{leung:DNA} or obtaining atomic resolution on the mica substrate\cite{fukuma:fm, hoogenboom:fm}. Nevertheless, the measured DNA height ($1.4\pm 0.1$~nm) is in excellent agreement with earlier low-force tapping-mode experiments with much softer (1~N/m) cantilevers\cite{MorenoHerrero:DNA}, though not as close to the 2~nm expected from the diameter of (uncompressed) double-stranded DNA as measured by tracking the resonance frequency of nanoscale cantilevers oscillating at subnanometer amplitudes\cite{leung:DNA}.

\section{\label{sec:comment}Conclusion}
In summary, we have prepared cantilevers with attached beads, that show quality factors $Q>20$ in water, yielding clearly recognizable resonances for different methods of cantilever actuation. This facilitates the application of dynamic-mode AFM in liquids and provides an alternative method to overcome the influence of noise in deflection detectors used for AFM\cite{hoogenboom:interferometer,fukuma:deflectiondetector}. It will also facilitate the application of frequency-modulation AFM in liquid\cite{fukuma:fm, hoogenboom:fm}: Though the frequency-modulation technique is sufficiently robust against slight detuning of the phase\cite{sader:fmafm}, its operation becomes highly impractical when the frequency feedback loop\cite{kahn:pll} can make discrete jumps between various spurious resonances close to the low-$Q$ cantilever resonance (see Fig.~\ref{fig:comparison}a, acoustic actuation). In spite of the orders-of-magnitude increase of the inertial mass of the cantilever, the ultimately achievable force sensitivity is hardly affected by the attachment of the bead to the cantilever. Optimum results are obtained for beads that do not significantly exceed the cantilever width, such that the viscous damping remains dominated by the cantilever shape and not by the attached bead. The cantilevers with attached beads can be well described by a simple mass-on-a-spring, damped harmonic-oscillator model, in which the effective mass is close to the bead mass.

In addition, we have demonstrated and extensively documented the attachment of beads and the implementation of a magnetic actuation system using a standard AFM combined with an inverted optical microscope. This will be of significant help for researchers interested in implementing these methods on any comined AFM/inverted-optical-microscope system.

We expect the results of this work to have broad applications for imaging soft, biological samples under near-physiological environments, and for accurate force measurements using dynamic-mode AFM in liquids.

\begin{acknowledgments}
We gratefully acknowledge Adam McKay for the DNA fragments, and Magnequench for a donation of the magnetic powder. This work was funded by the UK Biotechnology and Biological Sciences (BBSRC, BB/G011729/1 and BB/J006254/1) and Engineering and Physical Sciences (EPSRC, EP/K502959/1) Research Councils, and the US Office of Naval Research (N00014-10-1-0096).
\end{acknowledgments}

\appendix

\section{\label{sec:solenoid}Design and characterization of the solenoid}

\begin{figure}
\includegraphics{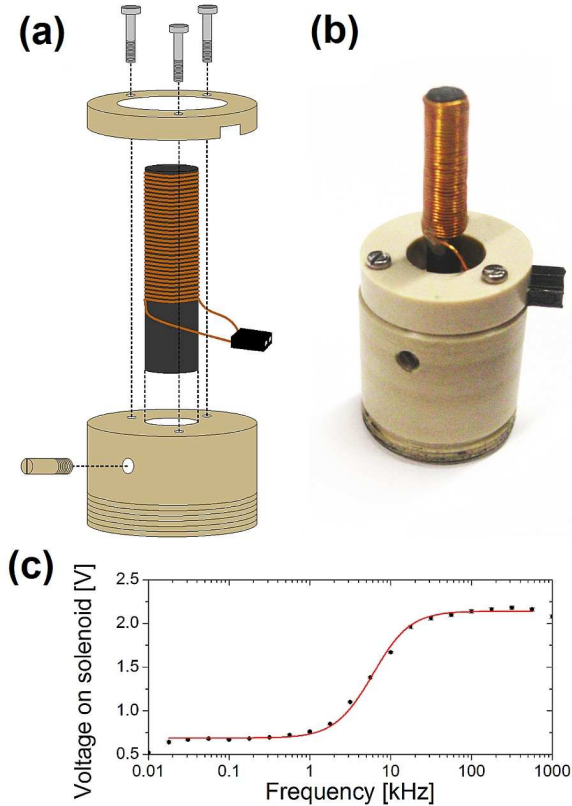}
\caption{\label{fig:coil} (Color online.) Design (a) and photograph (b) of the solenoid used for magnetic actuation. (c) The voltage (data points) over the solenoid as a function of frequency, as measured using a standard digital oscilloscope while holding the AC drive voltage constant at about 2~V. The solid line represents a fit to the data using Eq.~\ref{eq:solenoid_voltage}.}
\end{figure}

The solenoid for magnetic actuation was designed to fit into one of the objective threads of an inverted optical microscope (IX71, Olympus, Tokio, Japan). It can thus easily be adopted for any AFM system that is mounted on an inverted microscope platform. A sketch and a photograph of the solenoid can be seen in Fig.~\ref{fig:coil}a-b. It consists of a cylindrical ferrite rod (length = 30~mm, diameter = 4~mm, relative magnetic permeability $\mu_r$ = 31), 42 turns of copper wire (conductor diameter = 0.2 mm, insulation material: Polyimide, insulation thickness = 0.08 mm) and is fixed in a PEEK holder. It is connected to the external drive of the AFM with a BNC cable. In our experiments, a digital oscilloscope was used to monitor the voltage over the solenoid. 

\begin{figure}
\includegraphics{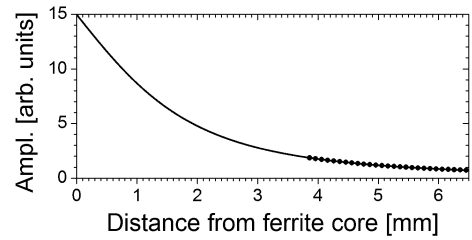}
\caption{\label{fig:BofZ} Distance dependence of the magnetic field generated by the solenoid, as measured via the amplitude of the magnetically actuated cantilever (dots, uncalibrated measurement). The cantilever with magnetic bead was immersed into a water filled petri dish such that the distance from the solenoid was a few mm. The solid line represents a fit with Eq.~\ref{eq:field}, by setting $R=2.18$~mm (based on the core diameter and the thickness of the wires) and $\ell = 15.5$~mm (as measured), and allowing a variable offset in $z$ (which defines the origin here).}
\end{figure}

The number of turns was optimized to generate a magnetic field in the frequency range of 10$\sim$100~kHz, corresponding to the range of cantilever resonances in this work. A larger number of turns would increase the magnetic field for lower frequencies, but for higher frequencies would increase the solenoid impedance to such values that the current (and thus the field) becomes limited by the maximum voltage (about $\pm 10$~V) of the excitation output of the AFM controller. This behavior was modeled by considering the equivalent electric circuit including the drive voltage ($V_d$), the resistance of the excitation output and cables connecting the coil ($R_c$), the Ohmic resistance ($R_s$) of the coil and its impendance due to its inductance $L_s$. The voltage over the solenoid ($V_s$) then follows from  
\begin{equation}
|V_s/V_d| =\frac{1}{\sqrt{1+\frac{2R_cR_s+R_c^2}{R_s^2+\omega^2L_s^2}}} \; .
\label{eq:solenoid_voltage}
\end{equation}
With a fixed $R_s=0.75\,\,\Omega$ (as measured independently), a fit to the measured $V_s$ (Fig.~\ref{fig:coil}c) yields $R_c=1.59\pm 0.04\,\,\Omega$ and $L_s=48\pm 2$~\textmu H, which is in agreement with the prediction $L_s=52\pm 3$~\textmu H from the expression\cite{albach:etechnik}
\begin{equation}
L = \mu_0\mu_r\frac{A}{\ell}N^2 \; ,
\label{eq:inductance}
\end{equation}   
where $A$ is the cross section of the solenoid, $\ell$ the length, $\mu_r$ the relative permeability, and $N$ the number of turns of the solenoid. Assuming a current $I$ and defining the central axis of the solenoid as the $z$-axis, and $z=0$ the end of the solenoid at which the field is determined ($z>0$), the field along the $z$ axis is given by
\begin{equation}
B_z = \mu_0 \mu_r\frac{ N I}{2\ell} \left (\frac{\ell+z}{\sqrt{R^2+(\ell+z)^2}}-\frac{z}{\sqrt{R^2+z^2}} \right ) \; ,
\label{eq:field}
\end{equation}
as derived by summing the fields of the individual current loops in the windings of the solenoid. For $\ell + z \gg R$, this results in
\begin{equation}
B_z(z) = \mu_0 \mu_r\frac{ N I}{2\ell} \left (1-\frac{z}{\sqrt{R^2+z^2}} \right ) \; .
\label{eq:approxfield}
\end{equation}
For the measurement range shown here, we did not find significant differences between the exact and the approximate equations. For magnetic actuation, the oscillation amplitude depends linearly on the magnetic field\cite{han:magnetic}. Fig.~\ref{fig:BofZ} shows the distance dependence of the oscillation amplitude in our setup. It demonstrates that the field decays relatively slowly on the length scale corresponding to practical solenoid-cantilever distances in an AFM system. Oscillation amplitudes of $\gtrsim 10$~nm were routinely achieved for cantilevers with attached magnetic beads. 

%

\end{document}